\documentclass[prl,aps,twocolumn]{revtex4}
\usepackage{graphicx}
\usepackage[abs]{overpic}
\usepackage{psfrag}
\usepackage{url}
\usepackage{amssymb,xcolor}

\begin{document}
\date{\today}
\title{When dunes move together, structure of deserts emerges}

\author{Mathieu G\'enois}
\author{Pascal Hersen}
\author{Sylvain Courrech du Pont}
\affiliation{Laboratoire Mati\`ere et Syst\`emes Complexes (MSC),
  Univ. Paris-Diderot, CNRS UMR 7057, 75205 Paris CEDEX 13}
\author{Guillaume Gr\'egoire}
\email{guillaume.gregoire@univ-paris-diderot.fr}
\affiliation{Laboratoire Mati\`ere et Syst\`emes Complexes (MSC),
  Univ. Paris-Diderot, CNRS UMR 7057, 75205 Paris CEDEX 13}
\affiliation{Academy of Bradylogists}

\begin{abstract}
  Crescent shaped barchan dunes are highly mobile dunes that are usually
  presented as a prototypical model of sand dunes. Although they
  have been theoretically shown to be unstable when considered
  separately, it is well known that they form large assemblies in
  desert. Collisions of dunes have been proposed as a mechanism to
  redistribute sand between dunes and prevent the formation of heavily large dunes, resulting in a
  stabilizing effect in the context of a dense barchan
  field. Yet, no models are able to explain the spatial structures of dunes observed in deserts.
  Here, we use an agent-based model with elementary
  rules of sand redistribution during collisions to access the full
  dynamics of very large barchan dune fields. Consequently, stationnary, out of equilibrium states emerge. Trigging the dune field density by a sand load/lost ratio, we show that large dune fields exhibit two assymtotic regimes: a dilute regime, where sand dune nucleation is needed
  to maintain a dune field, and a dense regime, where dune collisions
  allow to stabilize the whole dune field. In
  this dense regime, spatial structures form: the dune field is structured in narrow corridors of dunes extending in the wind direction, as observed in dense barchan deserts.\end{abstract}

\maketitle

In contrast with the layman's view, not all deserts are vast sand
seas. Depending on the variability of the local winds, sand dunes can
adopt various shapes. When viewed from above, they mimic large stars,
long linear ridges or crescent structures~\cite{Bagnold_1941,
  Pye_1990}. The crescent shaped dune, called
barchan~\cite{Bagnold_1941,Finkel_1959}, is a prototypical model of
sand dune dynamics~\cite{Kroy_2002_PRE} and its properties as an
isolated object are now well understood~\cite{Hersen_2004_EPJ,
  hersen_2002_PRL}. However, barchans are usually found in large dune
assembly, counting tens of thousands of
dunes~\cite{Elbelrhiti_2008_JGR, Bagnold_1941, Cooke_1993}. Barchan
fields are observed in regions where a rocky, non-erodible floor is
blown by a prevalent unidirectional flow and are
ubiquitous on Earth, on Mars and even underseas.   The very existence of a barchan field is in apparent contradiction with
the fact that barchan dynamic displays an unstable fixed point. Dunes
will either grow or shrink if their size departs from their
equilibrium size, set by the balance between sand loss and sand
capture. In contrast with this
unstable behavior, dune size ranges
from a few meters to several hundred of meters within a field. The size distribution does
not display a lack of small dunes or an anomalous number of huge
barchans~\cite{Elbelrhiti_2008_JGR, Duran_2009_GM,
  Duran_2011_NPG}. Furthermore, barchan fields may spatially be structured in narrow corridors, which extend in the wind direction. These corridors organize the dune field in stripes of
dense (resp. diluted) barchan areas, where dunes are smaller (resp.
larger), while neither the local conditions such as wind velocity or granulometry, neither the boundary conditions differ  ~\cite{Elbelrhiti_2008_JGR}. Thus, it is commonly thought that
dune-dune interaction, such as dune collisions, are at play to sustain a dune field over longtime, to structure the field in corridors and to select the dune size.  Field studies and underwater
experiments have shown that dunes can indeed exchange sands during
collision events~\cite{Vermeesch_2011_GRL, Endo_2004_GRL, Hersen_2005_GRL}. Dune collision
led to merge and split mechanisms, which can be stabilizing providing that large dunes are
regularly split into smaller ones, as proposed
using a mean field approach ~\cite{Hersen_2005_GRL}. Although this
idea is a first step, it is not enough to fully understand how
collisions set the structure and affect the stability of a large
assembly of dunes. Numerical studies implementing
phenomenological rules of collisions have been runned in order to forecast the statistical properties
of a large dune field in which collisions take
place~\cite{Duran_2009_GM}. If they succesfully recover a size selection, the critical aspect of spatial organization of dunes in corridors has never been taken in consideration yet. It is a general problem in which the large scale
emergent property comes from the complex, local, interactions of many
objects, what perfectly falls within the scope of an agent-based model.
\begin{figure}
  \begin{center}
    \includegraphics[width=0.4\textwidth, clip]{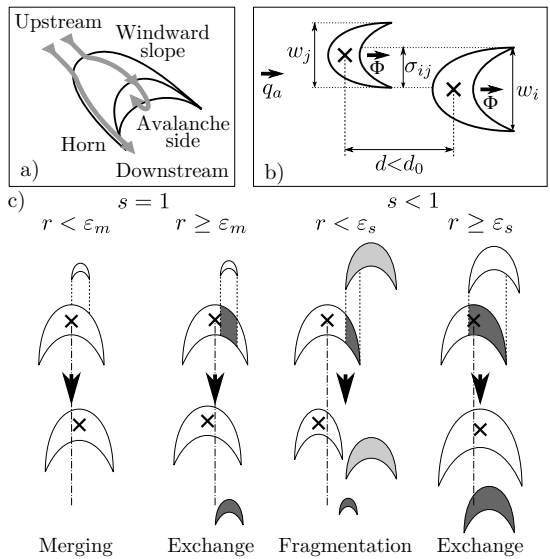}
    \caption{Elementary rules for barchan collisions (a) Morphology of
      a barchan. (b) Parameters
      used to describe a collision: two dunes of size $w_i$ and $w_j$
      interact when they are closer than $d_0$ and within a cross
      section of $\sigma _{ij}$. Each dune is losing a volume per
      unit of time $\Phi$ and it is fed by sand influx
      $q_a$. (c) The four cases of binary dune collisions. Depending on the upward projections $s=\frac{\sigma_{ij}}{w_j}$,
the collision is total ($s=1$) or lateral ($s<1$).
      Then, the downward projections $r=\frac{\sigma_{ij}}{w_i}$
      regarding to a merging $\varepsilon _m$ or a splitting
      $\varepsilon _s$ threshold is taken into account. The volume of sand is conserved during a collision. Grey levels show
      the sand redistribution. \label{Fcollisions}}
  \end{center}
\end{figure}

Here we use an agent-based model,
implementing dune collisions over the whole dune
field, to infer the dynamical statistical behavior of a large dune field in the limit of
long observation time, its stationary state and the possible emergence
of spatial structures. In order to identify the physical mechanisms involved in the emergent properties of dune fields, this model is restricted to the minimum ingredients of dunes dynamics and interactions.\\
The barchan shape is characterized by a low-slope upwind back and an
avalanche face downwind, which is framed by two arms pointing in the
wind direction (see Fig.~\ref{Fcollisions}(a-b)). The width, length
and height of barchans are linearly related to each others such that
their morphological state can be defined by one parameter only. Sand erosion and deposition processes force
barchans to move downwind. Their propagation velocity, which increases
with the wind shear stress and decreases with the sediment influx, is
inversely proportional to the dune size. The avalanche face acts as a
sand trap and barchans can propagate over long distances without
losing much sand. Yet, small sand loss occurs at the tip of the
barchan arms. Starting with a non-null
value, the sand loss increases very weakly with increasing dune
size, and can be considered as a constant~\cite{Hersen_2004_PRE}. On
the other hand, the input sand flux is proportional to the dune
width. As a result, the fixed point (where loss and gain are balanced)
is unstable and an isolated barchan can only grow or shrink and
eventually, disappear~\cite{Hersen_2004_PRE}. Indeed, below a critical
size, the barchan loses its avalanche face, turns into a dome-like
structure and quickly vanishes.

In our model, dunes are described by their width
$w$ only and, for the sake of simplicity, are cubic. They propagate
downwind at a speed $v$:
\begin{equation}\label{EqSpeed}
  v=\frac{\alpha}{w}.
\end{equation}
We assume that dunes lose sand homogeneously along their downwind
face. We call $\Phi$ the volume
lost per unit of time because of wind erosion. Barchans can also grow,
due to an incoming sand influx per unit of length transverse to the
wind, $q_{\mathrm{a}}$. The volume $V$ of an isolated dune will then
vary in function of time as:
\begin{equation}\label{EqVolume}
  \frac{dV}{dt}=-\Phi+q_{\mathrm{a}}w.
\end{equation}
Equation~\ref{EqVolume} contains the fundamental instability
of one isolated barchan of unstable equilibrium size $\tilde
w=\Phi/q_{\mathrm{a}}$. If the dune shrinks below the dome size $w_c$,
it is removed from the field. The model does not conserve the mass.  

One hypothesis to reconcile the unstable behavior of an isolated dune
(eq. \ref{EqVolume}) with the existence of dense barchan field is to
consider dune collisions. Smaller dunes are faster (eq. \ref{EqSpeed})
and can collide with larger, slower dunes what leads to a transfer of
mass between dunes ~\cite{Hersen_2005_GRL,Duran_2009_GM}.  Dune
collisions led to merge and split mechanism depending on the relative
size of dunes and their lateral alignment. Numerical
studies have shown that those parameters set the result of a
collision~\cite{Schwammle_2003_Nature,Katsuki_2005_JPSJ,
  Diniega_2010_Geom}. If the incoming dune is very small, it is simply
absorbed by the larger, slower one. If the dunes are of similar sizes,
a redistribution of mass occurs, and one (or several) small dunes are
emitted at the front while a larger dune is formed at the back.\\
In our model, two dunes are in interaction if they are
closer than a distance $d_0$ in the wind direction and if their
width projections overlap as shown in figure~\ref{Fcollisions}b. We
consider two types of interactions: a distant one through emitted sand
capture and sand flux screening and a close one through collision. The
distance $d_0$ reflects a typical distance for the sand flux to get
diluted laterally. Let's consider two dunes $i$ and
$j$, $i$ being the downwind dune and $j$ the upwind one. The size of
$i$ is noted $w_i$ and the overlapped width $\sigma_{ij}$. We define
the upward projection $s=\frac{\sigma_{ij}}{w_j}$ and the downward
projection $r=\frac{\sigma_{ij}}{w_i}$. The leeward dune catches a
part $s$ of the sand lost by the upwind dune. In the same time, the
upwind dune screens the leeward dune on the width $\sigma_{ij}$ from
any flux coming upwind of $i$ and $j$. If the upwind dune $j$ is the
only one that is closer than $d_0$ to the downwind dune $i$, the
volume of the latter varies as $d V_i / d t = s\times\Phi + q_a(w_i -
\sigma_{ij}) - \Phi$. Note that these eolian mass exchanges do not
affect the
aspect ratio of the dune (cubic) nor their position.\\
Two dunes collide when they overlap following the rules shown in
Fig.~\ref{Fcollisions}c. When $s\!=\!1$ (perfect overlap), $r$ is the
size ratio. The two dunes merge if $r$ is smaller than the merging
threshold $\varepsilon_{\mathrm{m}}$, and the new dune gets a volume:
$V_i^{t+\Delta t}=(w_i^{t})^3+(w_j^{t})^3$. When $s=1$ and $r
\geqslant \varepsilon_{\mathrm{m}}$, the total sand is redistributed
into two new dunes of volume: $V_i^{t+\Delta t} =
(w_i^{t})^3+(w_j^{t})^3 -\sigma _{ij}w_i^2$ and $V_j^{t+\Delta t} =
\sigma _{ij}w_i^2$. When $s<1$ (partial overlap), the sand is
redistributed into two or three dunes, respectively for $r$ values
bigger or smaller than the splitting threshold
$\varepsilon_{\mathrm{s}}$. When $s<1$ and
$r\geqslant\varepsilon_{\mathrm{s}}$, the two dunes exchange sand the
same way as when $s=1$ and $r\geqslant\varepsilon_{\mathrm{m}}$. When
$s<1$ and $r<\varepsilon_{\mathrm{s}}$, the bumping dune $i$ is
unaffected while the bumped dune is split in two dunes. The ejected
dune $k$ gets a volume: $V_k^{t+\Delta t} = \sigma _{ij}w_i^2$ while
$V_j^{t+\Delta t} = (w_j^{t})^3$ and $V_i^{t+\Delta t} =
(w_i^{t})^3-\sigma _{ij}w_i^2$. Note that the centers
of mass of the new dunes are set at the barycentric positions of the
incoming sand, which may shift the dunes laterally. To compensate for
sand dune loss, dunes can appear by nucleation anywhere in the dune
field where there is an empty place with a probability per unit time
and per unit of surface $\lambda$. Their size is arbitrary set to
$w_0$. These nucleations are the trace of topographical defects that
promotes sand deposition \cite{Pye_1990, Bagnold_1941}. This choice
maximizes the effect of noise. Thus any emerging behaviors will be
robust.
\begin{table}[t]
  \begin{center}
    \begin{tabular}{|*{12}{c|}}\hline
      $d_0$&$w_0$&$w_{\mathrm{c}}$&$\varepsilon_{\mathrm{s}}$&
      $\varepsilon_{\mathrm{m}}$&$\lambda
      ^{-1}$&$\alpha$&$\Phi\times10^{7}$
      &$q_{\mathrm{a}}$&$\ell$&$L$&$\Delta t$\\ \hline
      1&0.1&0.01&0.5&0.5&$2048$&$10^{-3}$&[1.5;500]&0&32&[32;128]&1\\ \hline
    \end{tabular}
    \caption{Parameters of the simulations.}\label{Tparameters}
  \end{center}
\end{table}

Looking for stationary behavior of large dune field, the dune field
boundaries are periodic and the field is long compared to the typical
distance of dune interactions, $L\gg d_0$. Since, dunes move along the
wind direction, we expect that the perpendicular direction $l$, does
not play a major role in dune-dune dynamics. The field is initially
filled with dunes at random positions homogeneously chosen. Their
sizes follow a constant probability which is centered on $w_0$ and
with a minimum cut-off value of $w_{\mathrm{c}}$. The numerical method
used to compute the assembly of dunes in this large field is based on
synchronous algorithm and off-lattice dynamics as in self-propelled
particles models~\cite{Chate_2008_EPJB}. By their non-trivial
kinematics, dunes can indeed be considered as self-propelled
particles, which exchange mass -- or momentum -- with their
neighborhood. Values of the different parameters used in the
simulations are reported in table~\ref{Tparameters}. In particular, we
assume that there is little sand around the dunes, so that the ambient
influx $q_{\mathrm{a}}$ is null. It implies that a lonely dune
always vanishes. From a simulation to an other, the dune density of
the field is trigged through a fixed nucleation rate $\lambda$ and a
changing erosion rate $\Phi$.

From the microscopic parameters, seven independent dimensionless
numbers can be built. One can define three length ratios
($w_{\rm{0}} / d_0$, $w_{\rm{c}} /d_0$, $\tilde w
/d_0=\Phi/(q_{\mathrm{a}}d_0)$---infinite here),
$\varepsilon_{\mathrm{s}}$ and $\varepsilon_{\mathrm{m}}$ which
control the dynamics of collisions and two times ratio. The three
relev	ant times are expected to be the time of disappearance of an
isolated dune of size $w_0$:
$t_{\mathrm{eol}}=(w_0^3-w_{\mathrm{c}}^3)/\Phi$, the typical
nucleation time: $t_{\mathrm{nuc}}=(\lambda\times d_0^2)^{-1}$ and a
collision time: $t_{\mathrm{col}}$, which could be evaluated as the
smallest time for two interacting dunes to collide
$t_{\mathrm{col}}=d_0/\left(\alpha/w_{\mathrm{c}}-\alpha/w_0\right)$. In
the present study, all parameters but $\Phi$ are kept constant (see table ~\ref{Tparameters}):
$t_{\mathrm{nuc}}=2048$ and $t_{\mathrm{col}}\simeq 11$ while
$t_{\mathrm{eol}}\in[20;6.7\,10^3]$, an isolated dune of size $w_0$
travels a distance $2$ to $660$ times its initial size before
disappearing. We define the control parameter $\xi$:
\begin{equation}\label{EqXi}
\xi=\frac{t_{\rm{eol}}}{t_{\rm{nuc}}}=\frac{w_0^3-w_{\mathrm{c}}^3}{\Phi}
\lambda d_0^2.
\end{equation} 
It measures the balance between disappearance of dunes due to loss of
sand and dune nucleation. So, one can
expect that a field gets emptied when $\xi\ll 1$.
\begin{figure}[t]
    \psfrag{N1}{$\scriptstyle N$}
    \psfrag{N2}{$\scriptstyle N$}
    \psfrag{t}{$t$}
    \psfrag{w}{$\left< w\right>$}
    \psfrag{w2}{$w$}
    \psfrag{P}{$P$}
    \psfrag{wc}{$w_c$}
    \psfrag{w0}{$w_0$}
    \psfrag{X1}{$\scriptstyle \xi=2.9$}
    \psfrag{X11}{$\scriptstyle\xi=2.9$}
    \psfrag{X2}{$\scriptstyle\xi=0.5$}
    \psfrag{X22}{$\scriptstyle\xi=0.5$}
    \includegraphics[width=0.4\textwidth,clip]{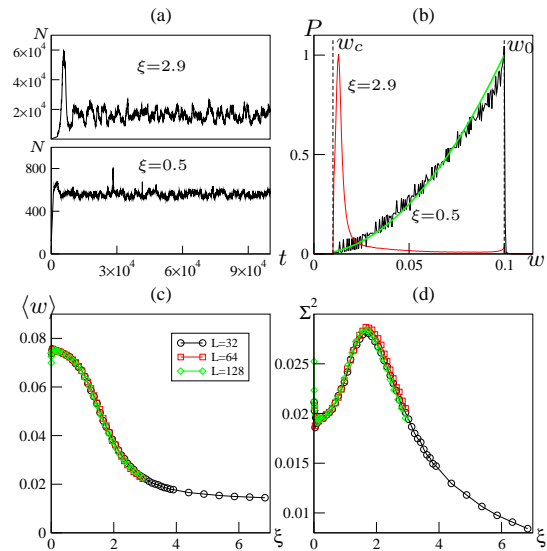}
    \caption{Characteristics of diluted and dense dune field. (a)
      number of dunes along time for dense and dilute regime. (b) size
      distributions for dilute and dense regime. The continuous green
      line is the analytic law of diluted distribution,
      eq.~\ref{Eqdistri}. (c) mean width $\left< w\right>$ and (d)
      variance of the width $\Sigma ^2$ \emph{versus} $\xi$ for
      different system lengths. Color online.\label{Fdistrib}}
\end{figure}

Interestingly, the dune field always reaches a stationary state within
our range of parameters (see Fig.~\ref{Fdistrib}a). For small $\xi$,
the barchan field is diluted, a few dunes are dispersed across the
whole field and dune collisions are
rare. Dunes can be considered as separate, unstable objects
whose disappearance is balanced by nucleation only. $\xi$ is
actually the exact dimensionless stationary density in the limit of
$\xi\ll1$. The normalized distribution of size $P(w)$ (see
Fig.~\ref{Fdistrib}b) follows the analytic distribution:
\begin{equation}\label{Eqdistri}
  P(w)=\frac{3w^2}{w_0^3-w_{\mathrm{c}}^3},\;\mbox{for}\;
  w\in[w_{\mathrm{c}};w_0],
\end{equation}
which can be derived from the individual dynamics
(Eq.~\ref{EqVolume}). Therefore, the typical size of dune is about
$3/4 \ w_0$, when $w_c$ is small enough. Such a field with low interaction between dunes compares to diluted deserts such as the barchan field of La Pampa de la Joya in Peru~\cite{Elbelrhiti_2008_JGR}.  

\begin{figure}
  \begin{center}
    \psfrag{Pn}{$P_n$}
    \psfrag{P}{$\scriptstyle P_n$}
    \psfrag{R}{$\scriptstyle \rho_{\mathrm{cl}}$}
    \psfrag{N1}{$N$}
    \psfrag{wm}{$\left<w\right>$}
    \includegraphics[clip,width=0.4\textwidth]{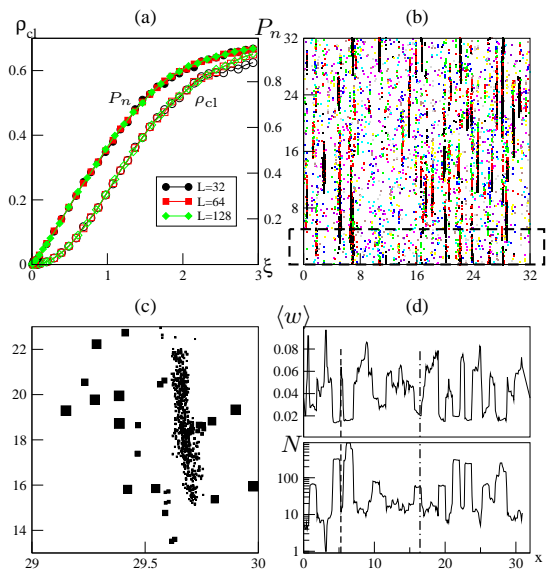}
    \caption{Spatial organization of a dune field (a) Density of
      clusters $\rho_{\mathrm{cl}}$ and probability $P_n$ for a dune
      to be in a cluster \emph{vs} $\xi$ for different system
      sizes. (b) snapshot of a computed desert at $\xi=2.9$, colors
      stand for dunes size. (c) closeup of a cluster of sub-figure
      (b). (d) Profiles of mean size $\left<w\right> $ and number $N$
      of dunes show anticorrelation, see \emph{e.g.} dashed and
      dotted-dashed lines. Profiles are computed by averaging in the
      bold dashed box shown on figure (b).(Color online. Movies are
      available as supplementary materials)\label{Fcluster}}
  \end{center}
\end{figure}

At large density $\xi\ge 1$ collisions dispatch sand in a non-trivial
manner. The field state remains stationary (Fig.~\ref{Fdistrib}a), but
the size distribution is fundamentally altered, and shifted to small
sizes (Fig.~\ref{Fdistrib}b). Dune collisions tend to increase the
number of dunes more rapidly than the effect of nucleation
itself. With $\varepsilon_{\mathrm{s}}$ and $\varepsilon_{\mathrm{m}}$
set to 0.5, the effective predominant collision type is the
fragmenting one that creates an additional dune
(Fig.~\ref{Fcollisions}c). Note that, the most probable size is close
but larger than the minimal size of dunes $w_{\mathrm{c}}$
(Fig.~\ref{Fdistrib}b). A dense assembly of dunes is not a trivial
homogeneous field with frequent collisions. On the contrary, dense
spatial clusters of small interacting dunes (\emph{i.e.} inter-dunes
is smaller than $d_0$) develop and gather the
major part of the dunes as seen on Fig.~\ref{Fcluster}. These cluster,
with sharp boundaries, are elongated in the wind direction, so that
the field is self-structured in a corridor-like pattern where the
local density is a highly fluctuating quantity. If we restrict our
measure to local low density, the dune size distribution shows a
maximum at $w_0$ as in a diluted desert: the local dune size is
directly correlated to the local density of dunes. A dense field looks
like a dilute field of big dunes with dense corridors of small dunes
(seef Fig.~\ref{Fcluster}d). These spatial structures and relation
between dune size and dilution are similar to what is observed in the
long barchan field that extends in the Atlantic Sahara (Morocco,
see~\cite{Elbelrhiti_2008_JGR} and supplementary material). Since many small
and fast dunes are created, another observed effect of collisions is
the spreading of the speed distribution. They can impact larger dunes
and lead to a succession of avalanche-like collisions. Thus,
transitory times are very different between diluted and dense
deserts. Whereas the number of dunes relaxes normally in diluted
deserts, dense deserts exhibit nearly periodic blow-up of their
population.

We identified two stationary states (at $\xi\ll 1$ and $\xi\ge 1$)
where the number of dunes but also the size distribution, the spatial
arrangement in the field and the relaxation to the equilibrium are dissimilar (fig. 2 and 3). This
is the signature of two different field dynamics: a diluted field
where dunes barely interact, and a dense field whose dynamics is
controlled by dune collisions. It is therefore legitimate to ask whether there is any phase transition when transiting from one to another. We checked that the mean
dune size and its variance change continuously when $\xi$ was varied
(Fig~\ref{Fdistrib}c and d). More generally, whatever the order
parameter we looked at, we found it to change smoothly without any
diverging moment. Furthermore, we did not detect any finite size
 effect (see Fig.~\ref{Fdistrib}c and~\ref{Fcluster}a), which
could sign a continuous~\cite{Privman_1990_book} or a first order
phase transition~\cite{Borgs_1990_JSP}. We neither found any influence
of initial conditions to the final state nor
detect any meta-stability. Even if the two
stationary states are very different, there is no phase transition but
rather a smooth cross-over when varying $\xi$ in this set of other
fixed parameters. Note that the system we studied is far
from equilibrium. Thus a phase transition was allowed even at low
dimension, in contrast with systems at
equilibrium~\cite{Mermin_1966_PRL}.

In conclusion, we introduced a minimal agent-based model of barchans
in interactions, in which kinematics and interactions are set in considering
experimental evidence of dune collisions. Its domain of validity
cannot extend outside the framework of asymptotic limits: infinite
size and infinite time of observations. However, varying the life time
of barchans due to sand loss, we showed a smooth cross-over between a
diluted desert to a dense desert where dunes aggregate in elongated
clusters. Computed deserts self-organize in
corridor-like patterns whith dense regions of small
dunes, and diluted spaces of larger dunes. This is observed in Earth dense barchan fields. Our model, although minimal, was able to capture the emergence of such heterogeneous patterning. In clusters, the typical barchan size is the result of avalanche of
collisions. Therefore, we demonstrated that a dune fragmentation mechanism, which seems to lack in previous studies, is a key process in setting the emergent properties of barchan dune fields. This fragmentation mechanism is here provided by collisions, but one could expect that other barchan destabilization mechanisms could play a similar role.


\end{document}